\documentclass[aps,prl,twocolumn,showpacs,superscriptaddress,groupedaddress]{revtex4}
\usepackage{graphicx}%
\usepackage{dcolumn}%
\usepackage{bm}%
\usepackage{amssymb,amsmath}   %
\usepackage{epsfig}
\usepackage{psfrag}
\usepackage{xspace}
\usepackage{multirow}
\usepackage{lscape}
\usepackage{comment}
\usepackage{color}

\newcommand{\dzero}     {D0}
\newcommand{\ttbar}     {\mbox{$t\bar{t}$}\xspace}

\newcommand{\ppbar}     {\mbox{$p\bar{p}$}\xspace}

\newcommand{\ljets}     {\mbox{$\ell$+jets}\xspace}

\newcommand{\mcatnlo}    {\mbox{\textsc{mc@nlo}}\xspace}
\newcommand{\geant}     {\mbox{\textsc{geant}}\xspace}

\newcommand{\met}{\mbox{\ensuremath{E\kern-0.6em\slash_T}}\xspace}
\newcommand{\metx}{\mbox{\ensuremath{E\kern-0.6em\slash_x}}\xspace}
\newcommand{\mety}{\mbox{\ensuremath{E\kern-0.6em\slash_y}}\xspace}
\newcommand{\sigmet}{\ensuremath{\sigma_{\mbox{\ensuremath{E\kern-0.6em\slash_T}}}}\xspace}

\newcommand{\ttfull}{\mbox{$t\bar{t} \rightarrow W^{+}bW^{-}\bar{b} \rightarrow \ell^{+} \nu b \ell^{-} \bar{\nu} \bar{b}$}}

\newcommand{\herwig}    {\mbox{\textsc{herwig}\xspace}}

\newcommand{\ttb}{\mbox{$t\bar{t}$}\xspace}

\newcommand{\qqbar}{\mbox{$q\bar{q}$}}

\newcommand{\llljcombi}{\ensuremath{11.8 \pm 3.2}\xspace}

\newcommand{\ljetsweight}{\ensuremath{64\%}\xspace}
\newcommand{\llweight}{\ensuremath{36\%}\xspace}

\newcommand{\m}{\phantom{$-$}}

\begin{document}

\hspace{5.2in} \mbox{Fermilab-Pub-12/319-E}

\title{\bf \boldmath Measurement of leptonic asymmetries and top quark polarization in $t\bar{t}$ production}

\date{June 21, 2012}

\begin{abstract}
We present measurements of lepton ($\ell$) angular distributions in
in top-quark ($t$) pair production and \ttfull\ decays produced in
\ppbar\ collisions at a
center-of-mass energy of $\sqrt{s}=1.96$~TeV, where $\ell$ is an
electron or muon. Using data corresponding to an integrated luminosity
of $5.4\text{~fb}^{-1}$, collected with the \dzero\ detector at the Fermilab
Tevatron Collider, 
we measure for the first time the leptonic
forward-backward asymmetry in dilepton final states and obtain
$A^{\ell}_{\rm FB}= (5.8 \pm 5.1 ({\rm
stat}) \pm 1.3 ({\rm syst}))$\%, corrected for detector acceptance.
This is compared to the standard model
prediction of $A^{\ell}_{\rm FB}({\rm predicted})=
(4.7\pm0.1)$\%. A deviation from the standard model prediction as
previously seen in a D\O\ measurement
based on the analysis of the $\ell +$jets final state is not observed in
these dilepton final states. The two results differ from each other by 1.4 standard
deviations and are 
combined to obtain
$A^{\ell}_{\rm FB}= (\llljcombi) \%$.  Furthermore, we present a first
study of the top-quark polarization.
\end{abstract}


\maketitle

To check the validity of the standard model (SM) of elementary
particle physics and to search for possible extensions, we measure the
properties of the top ($t$) quark. At the Tevatron \ppbar\ collider,
with $\sqrt{s}=1.96$~TeV, \ttb\ 
production is dominated by quark-antiquark (\qqbar) annihilation. At
leading order (LO) in 
perturbative quantum chromodynamics (QCD), production of \ttb\ pairs
through \qqbar\ annihilation is expected to be
forward-backward (FB) symmetric in the center-of-mass frame. At
next-to-leading order (NLO) QCD, interference leads to a small positive FB
asymmetry, which implies that the top (antitop) quark is emitted with
higher probability in the direction of the incoming quark
(antiquark) than in the opposite direction. 
Top pair production
through gluon-gluon fusion does not lead to such a FB asymmetry. 

SM predictions for the FB asymmetry
can be modified by processes beyond the
SM~\cite{Krohn:2011tw,AguilarSaavedra:2012ma}, such as contributions
from hypothesized axigluons~\cite{axigluons}, $Z'$ or $W'$
bosons~\cite{ZprimeWprime}, and new scalars~\cite{scalar}. 
These sources of physics beyond the SM also modify observables
sensitive to the top-quark polarization~\cite{Bernreuther:2010ny}.

The CDF and \dzero\ Collaborations have performed
measurements of the \ttb\ FB asymmetry in $\ell+\text{jets}$
final states involving signatures with exactly one charged lepton,
jets and an imbalance in transverse momentum
(\met)~\cite{Aaltonen:2011kc,Abazov:2011rq,Abazov:2007ab}. 
Both collaborations reported measured asymmetries
significantly larger than predicted in NLO QCD. 
\dzero\ finds a significant deviation from NLO QCD predictions of the
order of three standard deviations (SD)~\cite{Abazov:2011rq}. The
asymmetry in CDF data 
differs by more than three SD from the NLO QCD 
prediction at large values of the \ttb\ invariant mass ($m_t > 450$~GeV)
~\cite{Aaltonen:2011kc}. 
The ATLAS and CMS
Collaborations have performed measurements of the difference in
angular distributions between top quarks and antiquarks in the
$\ell+\text{jets}$ final state using asymmetries based on the top-quark
and antiquark rapidities~\cite{ATLAS:2012an,Chatrchyan:2011hk}
and pseudorapidities~\cite{Chatrchyan:2011hk}. The results are
consistent with the SM expectations. 

In this article, we test if an excess in the FB asymmetry similar to
the one seen in $\ell+\text{jets}$ final states can also be observed in 
previously unexplored dilepton final states, where the $W$ bosons 
from $t$ and $\bar{t}$ decays both decay into $e \nu_e$, $\mu \nu_\mu$, or
$\tau \nu_\tau$, and the $\tau$ lepton decays leptonically ($\tau
\to \ell \nu_\ell \nu_\tau$). 
We use data corresponding to an integrated luminosity of $5.4\text{~fb}^{-1}$, 
collected with the \dzero\ detector in Run~II of the Fermilab Tevatron Collider.

In addition to the \ttb\ FB asymmetry, where a full reconstruction of
the \ttb\ event is required, one can also study the FB asymmetry
through the $t$ and $\bar{t}$ decay products, for example, in the
distributions of charged leptons ($\ell=e,\mu$) from $t \to W^+b \to
\ell^+ \nu b$ and $\bar{t}
\to W^- \bar{b} \to \ell^- \bar{\nu} \bar{b}$ decays.
Here, we investigate such simpler leptonic
asymmetries. To study the nature of a
possible excess we  
present measurements of six different types of leptonic
asymmetries based on the
pseudorapidity and charge of the electrons or muons. Since these asymmetries are
determined from the angles of the charged leptons, 
they are measured with high resolution. In addition, we combine the
measurement of the leptonic
FB asymmetry with the \dzero\ measurement performed in 
$\ell+\text{jets}$ final states~\cite{Abazov:2011rq}. 
It is important to perform measurements of both the \ttb\ FB
asymmetry and the leptonic FB asymmetry, because their correlation
can be related through top-quark polarization to the underlying
dynamics of top-quark production~\cite{atfb_alfb}. For this reason, we
also present a first study of the longitudinal polarization of the top
quark in this article.

A description of the \dzero\ detector can be found in~\cite{d0det}.
The selection criteria and object identification of the dilepton
($ee,e\mu,\mu\mu$) decay channels follow those described in
Ref.~\cite{Abazov:2011cq}. To enrich the sample in
\ttbar\ events, we require two isolated, oppositely charged leptons
with transverse momentum $p_T>15$~GeV and at least two jets with
$p_T>20$~GeV and detector pseudorapidity
$|\eta_{\text{det}}|<2.5$~\cite{eta}. For the $e\mu$ channel we
require that $H_T$ (defined as the scalar sum of the larger of the two
lepton-$p_T$ values and the scalar $p_T$ of each of the two most
energetic jets) be greater than 110~GeV. For $ee$ and $\mu\mu$ events
we compute a likelihood for the significance of
\met~\cite{met_significance}, based on the probability distribution
calculated from the value of \met\ and the lepton and jet energy
resolutions. We require this likelihood to exceed the value typical for
background events. We find that only the $\mu\mu$ channel benefits
from an additional restriction on \met\ and, to increase signal
purity, we therefore require \met$>40$~GeV for the $\mu\mu$ final
state. We select a \ttb\ sample with a signal to background ratio of
3.2, 3.7 and 0.9 in the $ee$, $e\mu$ and $\mu\mu$ final states,
respectively. 

To simulate \ttb\ production, the \mcatnlo~\cite{mcatnlo}
generator is used assuming $m_{t}=172.5$~GeV. The
production of top quarks is simulated at NLO, while the decay is
simulated only at LO. To include full NLO QCD corrections to both
production and decay as well as mixed QCD and quantum electrodynamic
corrections and mixed QCD and weak corrections 
to the production amplitudes (denoted by ``QCD+EW''), we
simultaneously correct the normalized lepton and antilepton rapidity
distributions in \mcatnlo\ using the predictions of
Ref.~\cite{Bernreuther:2012ny}.  \herwig~\cite{herwig} is used to
simulate fragmentation, hadronization and 
decays of short-lived particles, and the generated events are
processed through a full detector simulation using
\geant~\cite{geant}.  The Monte Carlo (MC) events are overlaid with
data from random bunch crossings to model the effect of
detector noise and additional \ppbar\ interactions. The same
reconstruction programs are then applied to data and MC events. The
 background in the dilepton channel arises from
$Z/\gamma^{\ast}\rightarrow \ell^+\ell^-$ and diboson events ($WW$,
{\it WZ} and $ZZ$) with associated jets, from  instrumental 
background where a jet is misidentified as a lepton, and from  heavy
quarks that decay into leptons that pass isolation 
requirements.  A detailed description of these processes and their
generation can be found in Ref.~\cite{spin_paper}.

Leptons are reconstructed with  excellent resolution on the 
 measurements of their angles and electric charge. In contrast, it is challenging to
reconstruct the four-momenta of the $t$ and $\bar{t}$ quarks, since
the kinematics is underconstrained because of the two
neutrinos in the final state. Rather than determining the $t$ and
$\bar{t}$ four-momenta, as in
Refs.~\cite{Aaltonen:2011kc,Abazov:2011rq,Abazov:2007ab}, we measure
observables correlated to the FB asymmetry, which  depend
solely on the  $\eta$  and electric charge of the lepton $\ell$, as
proposed in Ref.~\cite{Bernreuther:2010ny}. The asymmetry for leptons
is defined as
\begin{equation}
\label{eq:afb_l}
A^{\ell} = \frac{N_{\ell^{+}}(\eta > 0) - N_{\ell^{-}}(\eta > 0)}{N_{\ell^{+}}(\eta > 0) + N_{\ell^{-}}(\eta > 0)}\\,
\end{equation}
where $N_{\ell^{-}}(\eta)$ and $N_{\ell^{+}}(\eta)$ correspond to the
number of leptons and antileptons as a function 
of $\eta$, respectively. If $CP$ invariance holds in \ttb\ production and decay, then
$N_{\ell^{+}}(\eta) = N_{\ell^{-}}(-\eta)$, and $A^{\ell}$ is equal to
the leptonic FB asymmetry, $A^{\ell} = A^{\ell^+}_{\rm FB} =
-A^{\ell^-}_{\rm FB}$ , defined as
\begin{equation}
\label{eq:afb_lplusminus}
A^{\ell^{\pm}}_{\rm FB} = \frac{N_{\ell^{\pm}}(\eta > 0) - N_{\ell^{\pm}}(\eta < 0)}{N_{\ell^{\pm}}(\eta > 0) + N_{\ell^{\pm}}(\eta < 0)}\\.
\end{equation}
The asymmetries $A^{\ell^+}_{\rm FB}$ and $A^{\ell^-}_{\rm FB}$ are
statistically independent and opposite. We can therefore combine the asymmetries
for $\ell^{+}$ and  $\ell^{-}$  by multiplying $\eta$ with the charge $Q$ of each lepton:
\begin{equation}
\label{eq:afb_lplusandminus}
A^{\ell}_{\rm FB} = \frac{N_{\ell}(Q\cdot\eta > 0) - N_{\ell}(Q\cdot\eta < 0)}{N_{\ell}(Q\cdot\eta > 0) + N_{\ell}(Q\cdot\eta < 0)}\\.
\end{equation}

In analogy to the FB asymmetry for $t$ and $\bar{t}$ quarks, we define
an angular asymmetry for leptons: 
\begin{equation}
\label{eq:afb_deltaY}
A^{\ell\ell} = \frac{N(\Delta\eta > 0) - N(\Delta\eta < 0)}{N(\Delta\eta > 0) + N(\Delta\eta < 0)},
\end{equation}
where $\Delta\eta=\eta_{\ell^+}-\eta_{\ell^-}$.
The asymmetry $A^{\ell}_{\rm CP}$ corresponds to a
longitudinal asymmetry in spin orientation relative to the
proton beam direction. It is defined as
\begin{equation}
\label{eq:afb_cp}
A^{\ell}_{\rm CP} = \frac{N_{\ell^+}(\eta>0) - N_{\ell^-}(\eta<0)}{N_{\ell^+}(\eta>0) + N_{\ell^-}(\eta<0)}\\.
\end{equation}
This asymmetry is sensitive to $s$-channel exchanges of heavy
nonscalar resonances with $CP$-violating couplings to quarks, but
not to possible $P$- and $CP$-violating effects from an
$s$-channel exchange of Higgs bosons~\cite{Bernreuther:2010ny}.

\begin{figure} %
\setlength{\unitlength}{1.1cm}
\begin{picture}(4.0,5.25)
\put(-2.9,2.20){\includegraphics[scale=0.27]{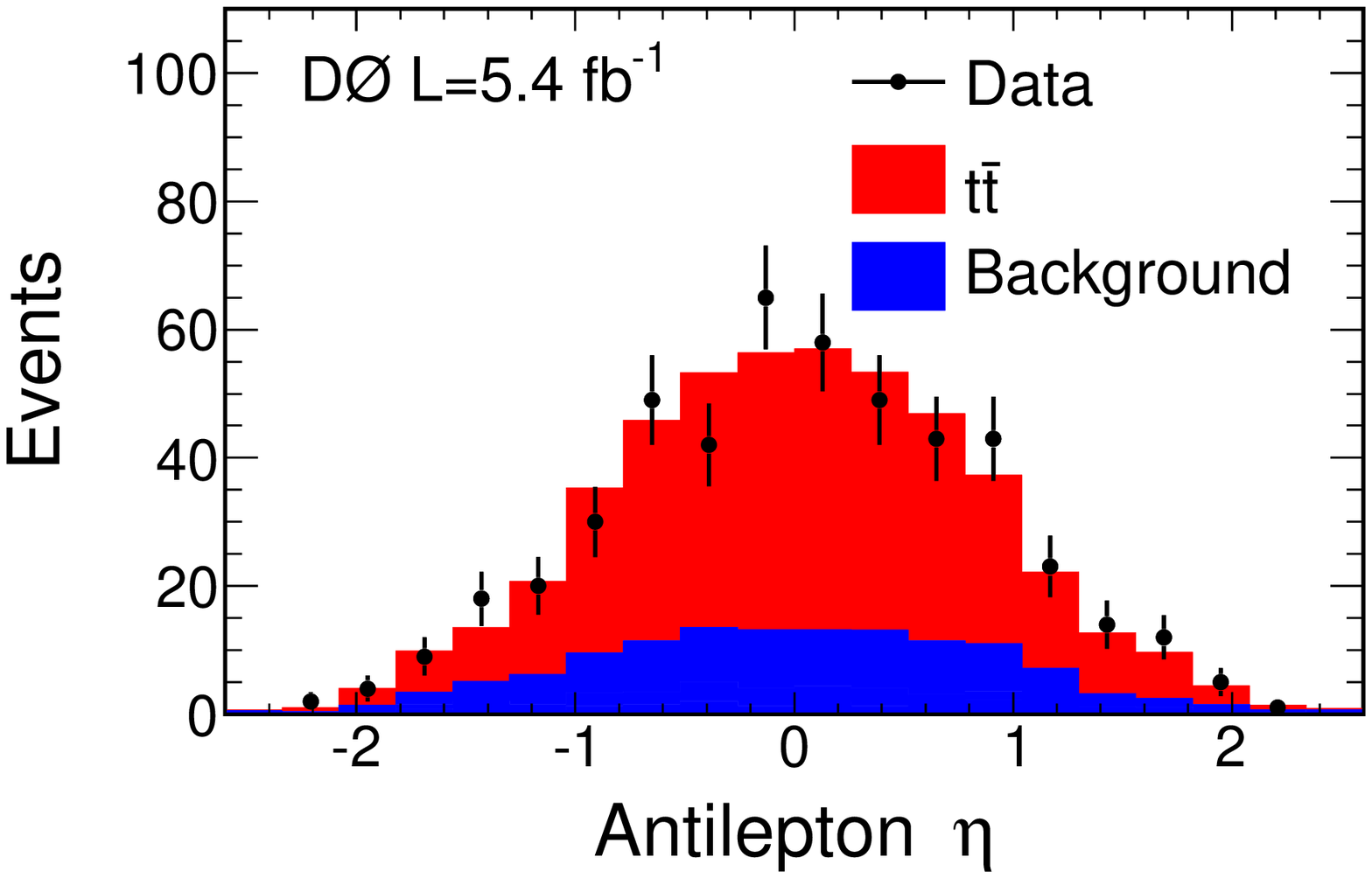}}
\put(1.70,2.20){\includegraphics[scale=0.27]{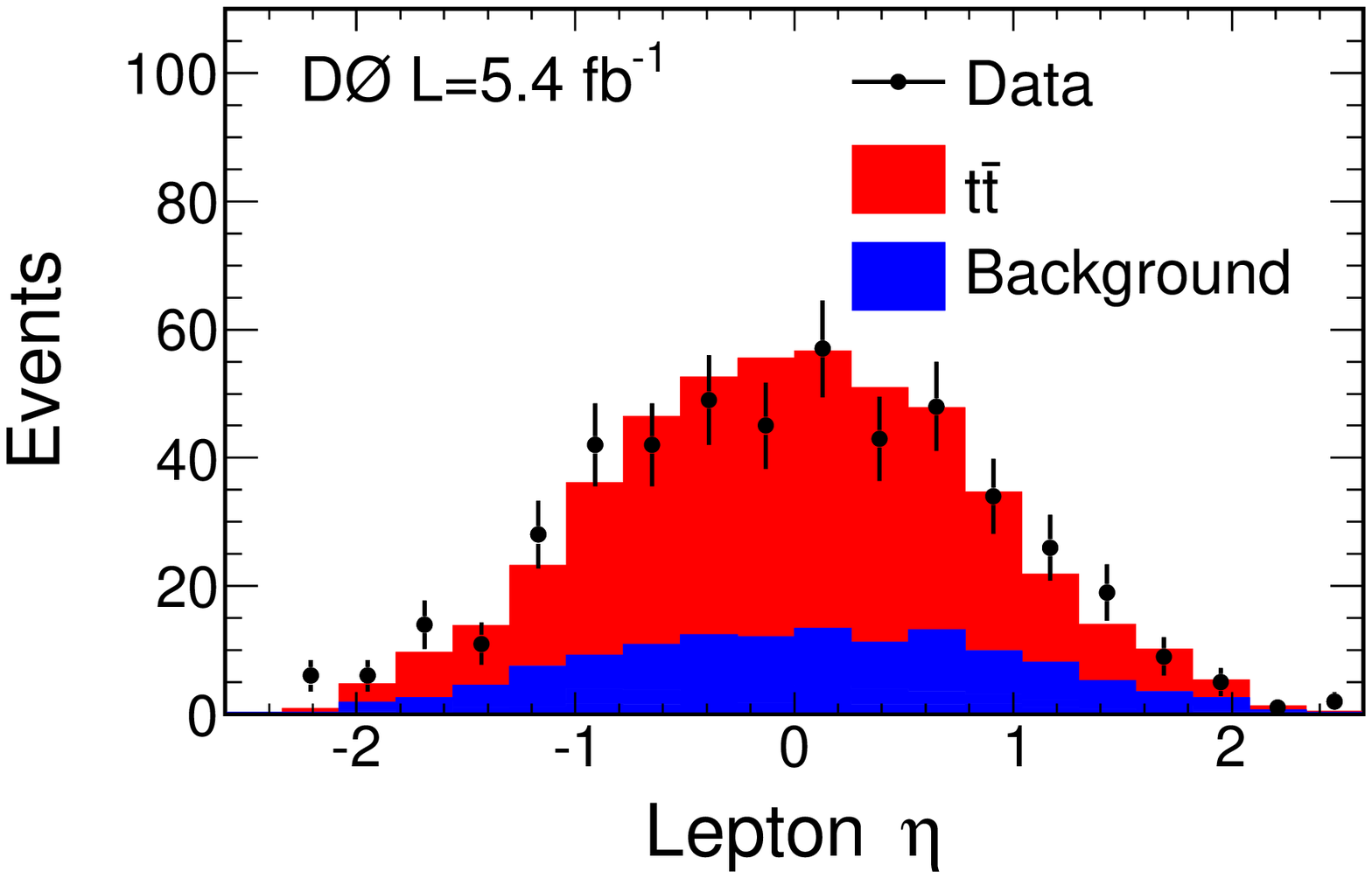}}
\put(-2.9,-0.9){\includegraphics[scale=0.27]{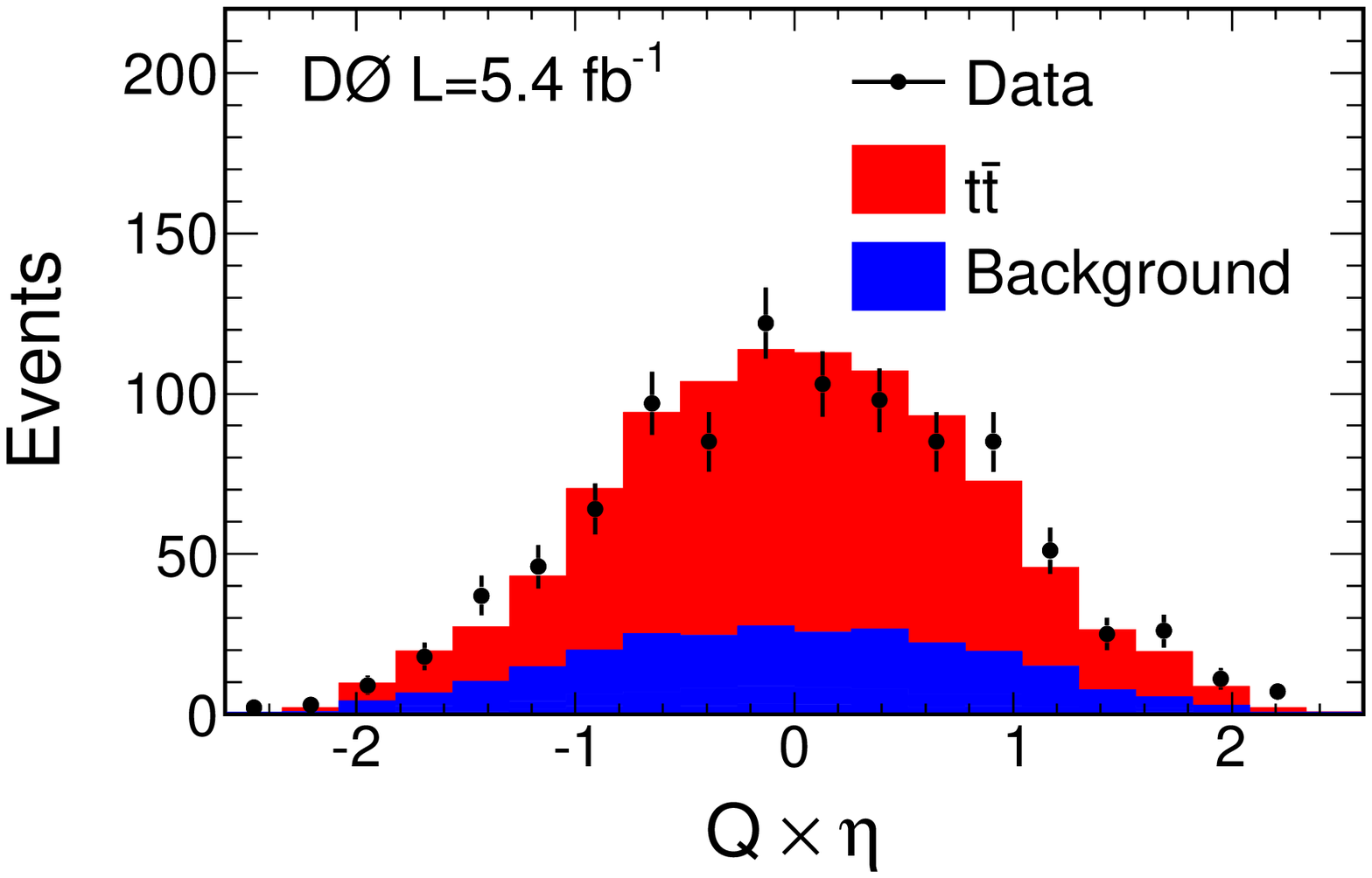}}
\put(1.70,-0.9){\includegraphics[scale=0.27]{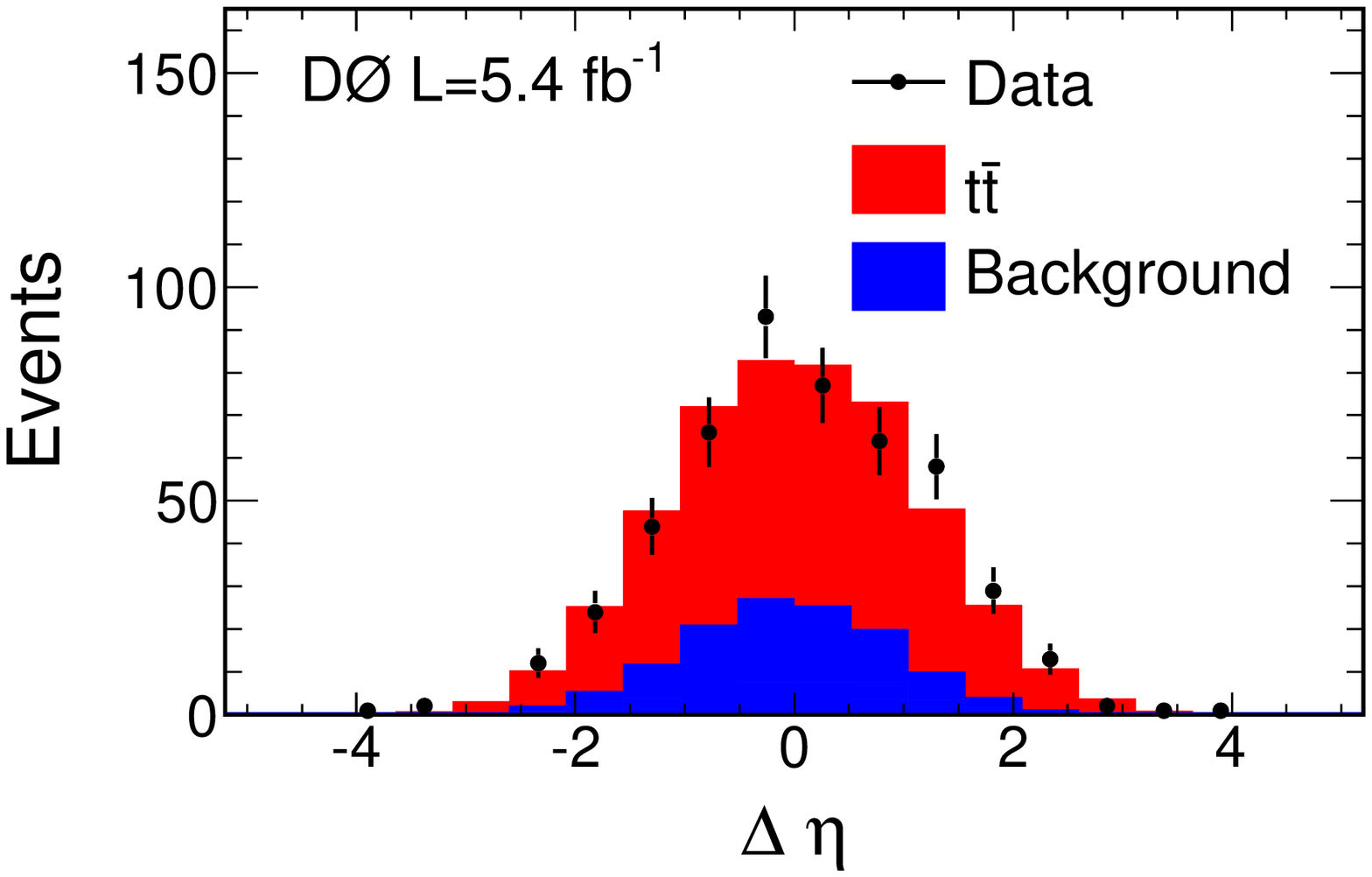}}
\put(-1.85, 4.2){\bf (a)}
\put( 2.75, 4.2){\bf (b)}
\put(-1.85, 1.15){\bf (c)}
\put( 2.75, 1.15){\bf (d)}
\end{picture}
\vspace{0.7cm}
\caption{\label{fig:lepton_rapidities} Pseudorapidity distributions
of the charged leptons for the combination of the $ee$, $e\mu$ and
$\mu\mu$ final states after the selection criteria have been
applied. The $\eta$ distribution of positively (a) and negatively (b)
charged leptons, the distribution of $Q\cdot\eta$ (c) and the
distribution of $\Delta \eta=\eta_{\ell^+}-\eta_{\ell^-}$ (d) are
shown. The vertical error bars indicate the statistical
uncertainty. The \ttb\ contribution is normalized to the data after
background subtraction.  }
\end{figure}

The asymmetries are measured in four ways using $\eta$ and $Q$ of the
leptons: separate $\eta$ distributions for (i) $\ell^+$ and (ii)
$\ell^-$, (iii) the charge-signed pseudorapidity, $Q\cdot\eta$, and
(iv) $\Delta\eta$. They are presented in
Fig.~\ref{fig:lepton_rapidities}.  To extract the asymmetries for
\ttb\ events from the distributions shown in
Fig.~\ref{fig:lepton_rapidities}, we subtract the background and then
correct for effects from event reconstruction and acceptance.  The
correction for detector acceptance is performed by multiplying the
background-subtracted number of events with the inverse of the
selection efficiency. This is calculated using $t\bar{t}$ MC events,
where we evaluate the selection efficiency separately for 20 bins
in lepton $\eta$, to reduce the model dependence of our acceptance
correction and to provide sufficient MC statistics.

The resolution of the measurement of lepton $\eta$  is obtained
from studies of \ttb\ MC events by comparing the generated value of
$\eta$ with the value measured following the event reconstruction.  For
electrons and muons, we use the $\eta$ of tracks measured in the
tracking system and find this resolution to be the same for
both types of leptons. This resolution is also investigated using cosmic-ray
muons that appear as dimuon events and is found to be $\approx
0.0026$, consistent with the MC expectation. For $\approx 99.8$\% of
the electrons or muons in \ttb\ MC events, the
sign of lepton $\eta$ is correctly reconstructed. Migration of
events within the ``forward'' or ``backward'' regions does not
affect the reconstructed angular asymmetry except for negligible acceptance
corrections. The
reconstruction effects on the measurement of $\eta$ can therefore be
neglected for charged leptons.

The $Z$+jets background, which is predicted through MC
simulation~\cite{spin_paper}, contributes to the asymmetry. To study
the influence of the $Z$+jets background, we perform measurements of
all six asymmetries in a sample dominated by $Z$+jets production in
final states with two electrons or two muons.  Applying the same event
selections as for the final \ttb\-enriched sample, except for the
\met\ significance likelihood and \met\ requirements, all asymmetries
are measured using the same procedure as for the measurement of \ttb\
asymmetries, but treating $Z$+jets as ``signal'' and $t\bar{t}$ as
``background.'' In this control sample, all other background
contributions are negligible.  The data and MC predictions for the
$\eta$ distribution of positively and negatively charged leptons, for
$Q\cdot\eta$, and $\Delta
\eta$, are in good agreement, as presented in~\cite{epaps}.

To verify that the measurement of the \ttb\ asymmetries is unbiased
and correctly estimates the statistical uncertainty of the result, we
perform the measurement using ensembles of MC pseudoexperiments. To
obtain samples with different asymmetries, we mix a \ttb\ MC event
sample weighted to have no asymmetry with different fractions of \ttb\
MC events with a SM asymmetry. We fluctuate the expected number of
events in the ``forward'' and ``backward'' direction for each
pseudoexperiment assuming Poisson statistics and apply the same
procedure as for data to extract the asymmetry. This test shows that
the measurement is unbiased and that the statistical uncertainties are
estimated correctly.

Systematic uncertainties can affect the distributions in lepton
$\eta$. In particular, the energy scale for jets, the jet
energy re\-so\-lu\-tion, the jet reconstruction, the normalization of background, 
the MC-derived acceptance, and the finite number of MC events can 
shift the measured asymmetry.
 The normalization of the background has uncertainties from diboson 
and $Z$+jets cross sections, as well as a $6.1\%$ uncertainty
on the data sample's integrated luminosity.  The systematic
uncertainties on the light- and heavy-flavor jet energy scales,
the jet energy re\-so\-lu\-tion, and the jet reconstruction can affect the
acceptance. We evaluate the size of these uncertainties by applying the
variation in acceptance corrections and in the differential distribution of lepton
$\eta$  in deriving the \ttb\ asymmetry.

In addition, we compare the acceptance from single leptons obtained
from simulated  \ttb\ events with the acceptance obtained from $Z \rightarrow
\ell^{+}\ell^{-}$ data.  We select a data sample enriched
in $Z \rightarrow \ell^{+}\ell^{-}$ events, where one lepton is
required to pass tight lepton-selection criteria to function as a ``tag''
and the other ``probe'' lepton to pass a loose lepton selection. The
acceptance is evaluated as a function of $\eta$ by applying a
tight-lepton identification requirement on the probe.  No significant
difference is observed between the acceptance for positive or negative
pseudorapidities, or between positively and negatively charged
leptons.  
A systematic uncertainty on the acceptance is defined for each lepton
charge by the difference in acceptance between the forward and
backward hemisphere of the detector.
This study is performed separately for electrons and
muons. The systematic uncertainties are added in quadrature to yield
the total systematic uncertainties given in
Table~\ref{tab:combo_systematics_afb}.

\begin{table}
\caption{\label{tab:combo_systematics_afb}
Systematic uncertainties for the six unfolded asymmetries defined in
Eqs.~(\ref{eq:afb_l})--(\ref{eq:afb_cp}) for the combination
of all dilepton final states. All values are given in \%.}
\begin{tabular}{l|cccccc}
\hline
\hline
Source             & $A^{\ell}$ & $A^{\ell^{+}}_{\mathrm{FB}}$ & $A^{\ell^{-}}_{\mathrm{FB}}$ & $A^{\ell}_{\mathrm{FB}}$ & $A^{\ell\ell}$ & $A^{\ell}_{\mathrm{CP}}$\\
\hline
Jets               & $1.1$ & $0.8$ & $1.7$ & $1.0$ & $1.5$ & $1.2$\\
MC statistics     & $0.4$ & $0.4$ & $0.4$ & $0.3$ & $0.5$ & $0.3$\\
Background normalization & $0.3$ & $0.3$ & $0.6$ & $0.3$ & $0.7$ & $0.3$\\
Acceptance        & $0.7$& $0.2$ & $1.5$ & $0.7$ & $2.3$ & $0.9$\\
\hline
Total             & $1.4$ & $1.1$ & $2.4$ & $1.3$ & $2.9$ & $1.6$\\
\hline
\hline
\end{tabular}
\end{table}

Using the distributions in Fig.~\ref{fig:lepton_rapidities}, the lepton
asymmetries of Eqs.~(\ref{eq:afb_l})--(\ref{eq:afb_cp}) are measured. 
The raw asymmetries are corrected for acceptance effects
(``unfolded'') and compared to the predictions from \mcatnlo\ including QCD+EW
corrections~\cite{Bernreuther:2010ny}. All values are listed in
Table~\ref{tab:signal_asymmetries}. The unfolded asymmetries are in
agreement with the SM predictions within errors.
\begin{table}
\caption{\label{tab:signal_asymmetries}
Measured asymmetries for leptons, as
  defined in Eqs.~(\ref{eq:afb_l})--(\ref{eq:afb_cp}), including statistical
and systematic uncertainties for the combined dilepton final states
using raw and unfolded distributions
are compared to predictions
from \mcatnlo\ including QCD+EW corrections. Our predictions are
calculated using the NLO QCD+EW distributions in both numerator and
denominator of Eqs.~(\ref{eq:afb_l})--(\ref{eq:afb_cp}). This is
different from the calculations in Refs.~\cite{Bernreuther:2010ny,Bernreuther:2012ny} where the denominator is calculated in LO QCD to
derive expressions for the asymmetries of ${\cal O}(\alpha_s)$.
All values are given in \%.
}
\begin{tabular}{l|ccc}
\hline
\hline
                        &  Raw   &  Unfolded & \m Predicted \\
\hline
$A^{\ell}$              & \m$ 2.9 \pm 6.1 \pm 0.9 $ & \m$ 2.5 \pm 7.1 \pm 1.4 $ & \m$4.7 \pm 0.1$\\
$A^{\ell^{+}}_{\rm FB}$ & \m$ 4.5 \pm 6.1 \pm 1.1 $ & \m$ 4.1 \pm 6.8 \pm 1.1 $ & \m$4.4 \pm 0.2$\\
$A^{\ell^{-}}_{\rm FB}$ &   $-1.2 \pm 6.1 \pm 1.3 $ &   $-8.4 \pm 7.4 \pm 2.4 $ &  $-5.0 \pm 0.2$\\
$A^{\ell}_{\rm FB}$     & \m$ 3.1 \pm 4.3 \pm 0.8 $ & \m$ 5.8 \pm 5.1 \pm 1.3 $ & \m$4.7 \pm 0.1$\\
$A^{\ell\ell}$          & \m$ 3.3 \pm 6.0 \pm 1.1 $ & \m$ 5.3 \pm 7.9 \pm 2.9 $ & \m$6.2 \pm 0.2$\\
$A^{\ell}_{\rm CP}$     & \m$ 1.8 \pm 4.3 \pm 1.0 $ &   $-1.8 \pm 5.1 \pm 1.6 $ &  $-0.3 \pm 0.1$\\
\hline
\hline
\end{tabular}
\end{table}
The asymmetry $A^{\ell}_{\rm FB}$ defined in
Eq.~(\ref{eq:afb_lplusminus}) is also measured in \ljets\ final
states~\cite{Abazov:2011rq}. 
The result for $A^{\ell}_{\rm FB}=(15.2 \pm 4.0)$\% is compared to a
predicted value from \mcatnlo\ of $(2.1 \pm 0.1)\%$ in Ref.~\cite{mcatnlo}. 
We checked that our current predicted asymmetry of $(4.7 \pm 0.1)\%$
is independent of the final state and that the difference from the
prediction given in Ref.~\cite{mcatnlo}, which should be considered superseded,
is only due to the additional QCD+EW corrections as described
previously. The dominant
systematic uncertainty on the prediction and on our measurement in
dilepton final states is given by jet reconstruction related
systematics. The total uncertainty of the measurement is dominated by
the statistical component.  

Since the \ljets\ and dilepton final
states are selected to be statistically independent, we can combine
the two asymmetries $A^{\ell}_{\rm FB}$ 
using the BLUE method~\cite{blue:lyons,blue:valassi}. All systematic 
uncertainties evaluated in both measurements are treated as fully correlated. 
The difference between the two individual measurements is 1.4 SD.
The combination yields a leptonic FB asymmetry of
$A^{\ell}_{\rm FB} =(\llljcombi)$\%, where the \ljets\  
channel contributes \ljetsweight and the dilepton channel
\llweight of the information. This represents 
an improvement of about $20$\% relative to the 
uncertainty in the \ljets\ channel alone.
Comparing the combined result to the predicted leptonic FB
asymmetry from \mcatnlo\ plus higher order QCD+EW corrections,
$A^{\ell}_{\rm FB}({\rm predicted}) = (4.7 \pm 0.1)$\%, we observe  
a disagreement at the level of 2.2 SD.

To further investigate this deviation of the asymmetry from the SM
prediction, we analyze the longitudinal polarization of the top quark.
While in the SM top quarks are expected to be produced unpolarized in
\ttb\ events, there are many beyond the SM models that would enhance
the \ttbar\ FB asymmetry~\cite{Krohn:2011tw} and therefore the
leptonic asymmetries defined in
Eqs.~(\ref{eq:afb_l})--(\ref{eq:afb_cp}), and would also lead to a
nonvanishing longitudinal polarization of the top quark. Examples are
models with new parity-violating interactions.  In the absence of
effects from acceptance, the distribution of $\cos\theta^-$ and
$\cos\theta^+$ should be isotropic~\cite{Bernreuther:2010ny} for
unpolarized top quarks, where $\theta^+$ ($\theta^-$) is the angle
between the direction of the $\ell^+$ ($\ell^-$) in the $t$
($\bar{t}$) rest frame and the $t$ ($\bar{t}$) direction in the
\ttbar\ rest frame. A longitudinal polarization of the top quark would
cause asymmetric $\cos\theta^{\pm}$ distributions.

Assuming $CP$ invariance, i.e. that the distributions of $\cos\theta^+$ and
$\cos\theta^-$ are equal, we measure the distribution f $\cos\theta$,
defined by the sum of the $\cos\theta^{\pm}$ distributions.
The
calculation of the angles $\theta^{\pm}$ requires a transformation of
the momenta of the charged leptons into the $t$ and $\bar{t}$ quark
rest frames. Every event must therefore be fully reconstructed. This
is performed using the neutrino weighting method, devised originally
to measure the top-quark mass in the dilepton channel \cite{nuWtmass}
and recently applied to measure $t\bar{t}$ spin
correlations~\cite{spin_paper}.

\begin{figure} %
\setlength{\unitlength}{1.1cm}
\begin{picture}(4.0,2.5)
\put(-2.95,-0.25){\includegraphics[scale=0.27]{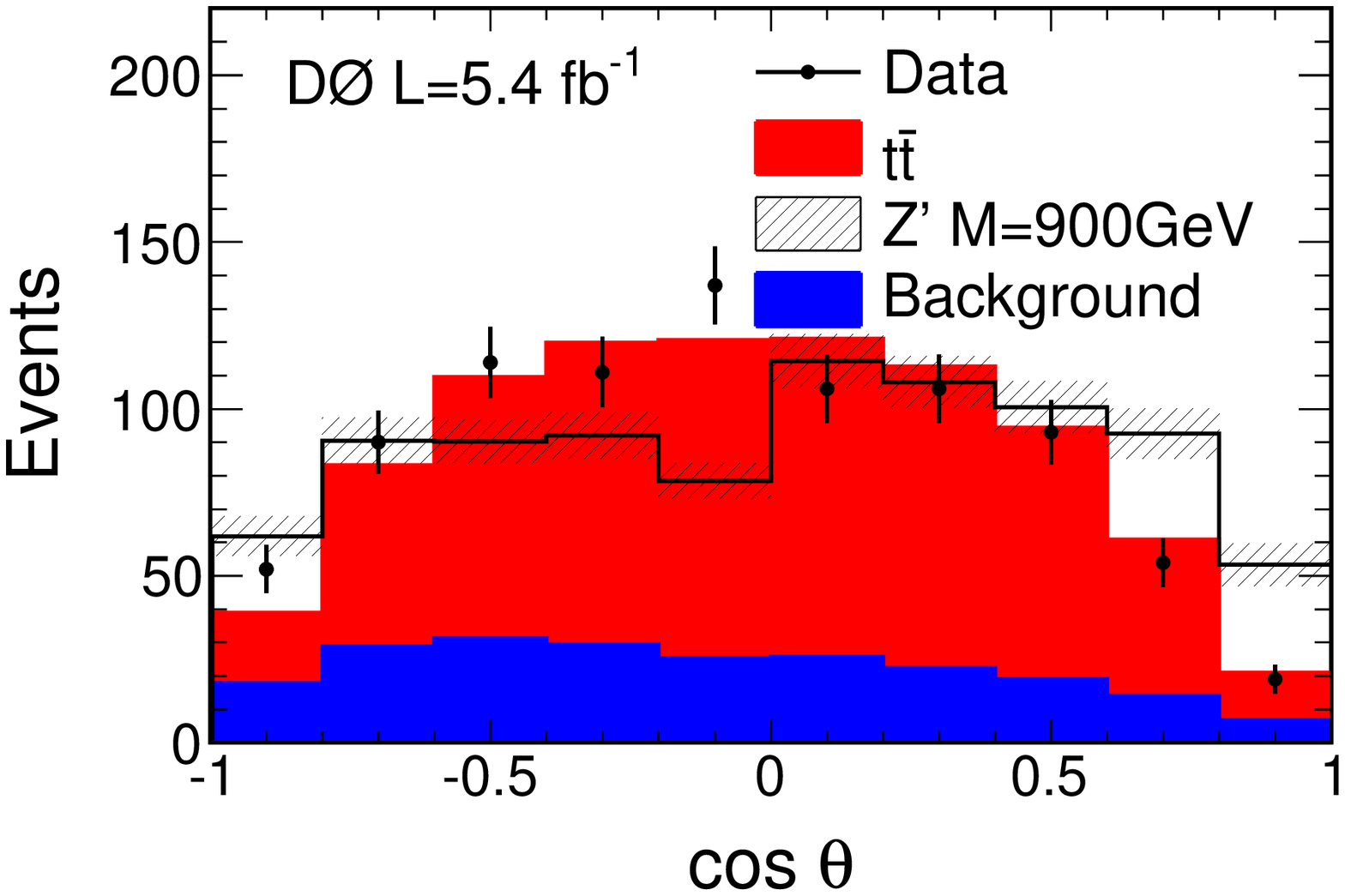}}
\put( 1.7, -0.25){\includegraphics[scale=0.27]{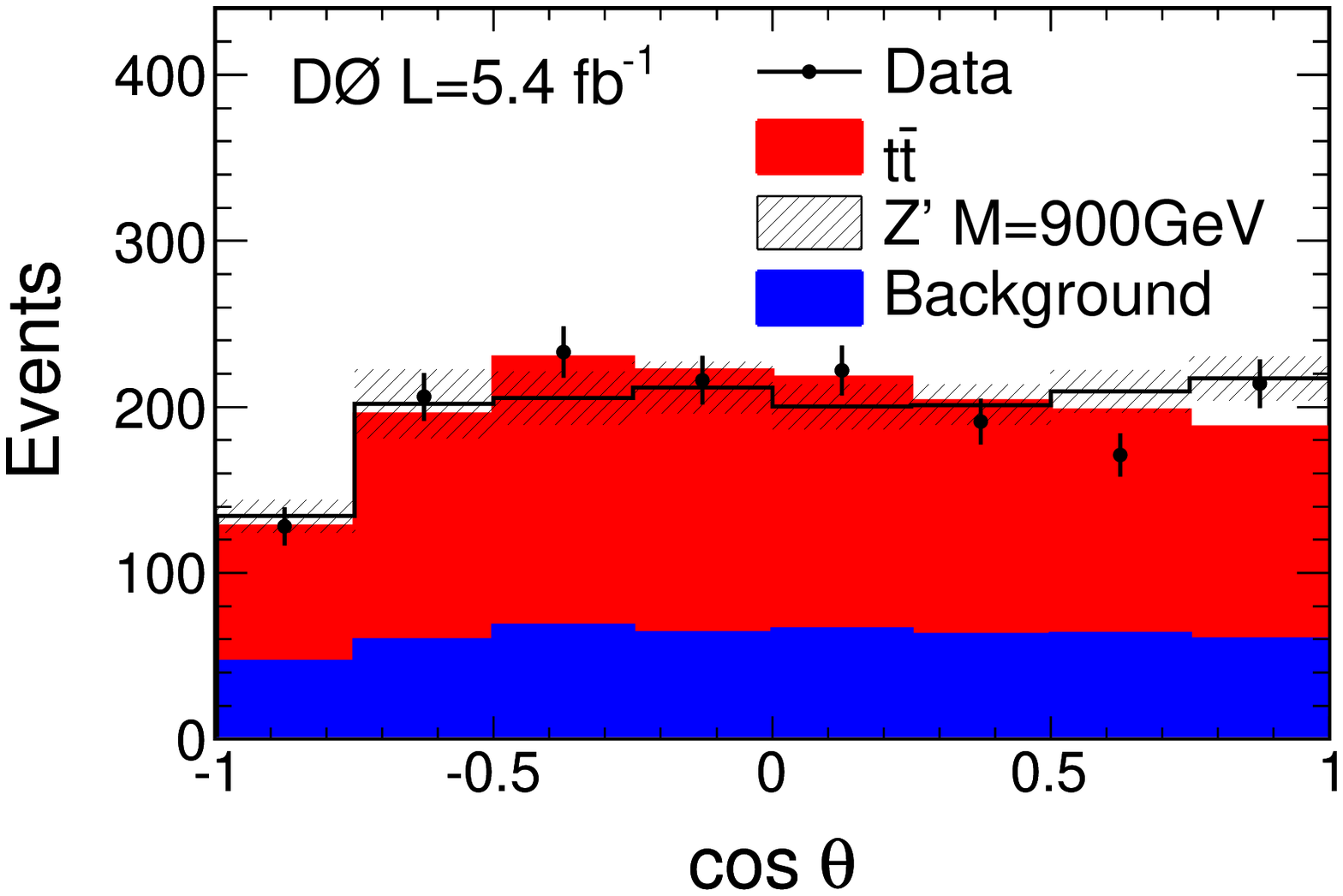}}
\put(-1.85, 2.10){\bf (a)}
\put( 2.70, 2.10){\bf (b)}
\end{picture}
\caption{\label{fig:cos_distributions} The distribution of
$\cos \theta$ is shown for the combination of the dilepton channels
(a) and the $\ell$+jets channels (b). The data are compared to the SM
predictions. The vertical error bars on the data points indicate the
statistical uncertainty of the data. The distribution of \ttb\ pairs
produced via a hypothetical $Z'$ boson is also shown; the uncertainty
due to the limited size of the MC sample is shown by the shaded
band. The same $Z'$ model as in \cite{Harris:1999ya,Abazov:2011gv} is
used.}
\end{figure}

In Fig.~\ref{fig:cos_distributions}, the $\cos\theta$ distribution is
shown separately for the dilepton and \ljets\ final states. The distribution 
for \ttb\ events produced via a leptophobic topcolor $Z'$ boson,
with the same parity-violating couplings to quarks as the SM $Z$ boson and 
a width $\Gamma = 0.012 M_{{\mathrm Z}}$
\cite{Harris:1999ya,Abazov:2011gv} is also shown to illustrate the
effect of producing $t$ and $\bar{t}$ quarks with longitudinal polarization.
The agreement between the data and the SM prediction in both
distributions is good, yielding a Kolmogorov-Smirnov test probability
of $14$\% in the dilepton channel and $58$\% in the $\ell$+jets
channel. There is no significant hint of new sources of
parity violation leading to a longitudinal polarization in \ttb\
production.

In conclusion, we measured angular asymmetries in \ttb\ production
based on $\eta$ distributions of charged leptons in dilepton final
states for the first time.
We find the leptonic FB asymmetry $A^{\ell}_{\rm FB}$ and the lepton
asymmetry $A^{\ell\ell}$ in agreement with zero and with the SM prediction
and do not observe an excess such as previously reported in \ljets\
final states. Combining our measurement of $A^{\ell}_{\rm
FB}$ with the measurement performed using leptons in \ljets\ final
states yields $A^{\ell}_{\rm FB}=(\llljcombi)$\%, which is 2.2 SD
above the calculated value including higher order QCD+EW corrections
of $A^{\ell}_{\rm FB}({\rm 
predicted}) = (4.7 \pm 0.1)$\%.  To explore the nature of this deviation,
the top-quark polarization in the dilepton and
$\ell+$jets final states has been studied for the first time and shows
good agreement between the data and the SM prediction in both
channels.

\section*{Acknowledgments}
We would like to thank W. Bernreuther and  Z.~G.~Si for useful
discussions and providing us
with the lepton rapidity distributions used for theoretical predictions. 
%
We thank the staffs at Fermilab and collaborating institutions,
and acknowledge support from the
DOE and NSF (USA);
CEA and CNRS/IN2P3 (France);
FASI, Rosatom and RFBR (Russia);
CNPq, FAPERJ, FAPESP and FUNDUNESP (Brazil);
DAE and DST (India);
Colciencias (Colombia);
CONACyT (Mexico);
NRF (Korea);
CONICET and UBACyT (Argentina);
FOM (The Netherlands);
STFC and the Royal Society (United Kingdom);
MSMT and GACR (Czech Republic);
BMBF and DFG (Germany);
SFI (Ireland);
The Swedish Research Council (Sweden);
and
CAS and CNSF (China).

\newpage


\begin{thebibliography}{99}
\bibitem{eta} The rapidity $y$ and pseudorapidity $\eta$ of a particle are defined as functions
    of the polar angle $\theta$ with respect to the proton beam
    direction and velocity $\beta$ as
    $y(\theta,\beta) \equiv {\frac{1}{2}}
    \ln{[(1+\beta\cos{\theta})/(1-\beta\cos{\theta})]}$ and
    $\eta(\theta) \equiv y(\theta,1)$, where $\beta$ is the ratio of a
    particle's momentum to its energy. We distinguish detector $\eta$
    ($\eta_{\text{det}}$) and collision $\eta$, where the former is defined
    with respect to the center of the detector and the latter with
    respect to the \ppbar\ interaction vertex.
\end{thebibliography}
\end{document}